\documentclass[sigconf]{acmart}

\usepackage{amsmath,amsfonts}
\usepackage{algorithmic}
\usepackage{algorithm}
\usepackage{array}
\usepackage[caption=false,font=normalsize,labelfont=sf,textfont=sf]{subfig}
\usepackage{stfloats}
\usepackage{url}
\usepackage{verbatim}
\usepackage{graphicx}
\usepackage{hyperref}
\usepackage{xurl}

\usepackage{multirow}
\usepackage{tabularx}
\usepackage{booktabs}
\usepackage{color}
\usepackage{threeparttable}
\usepackage{subfig}
\usepackage{enumitem}
\usepackage{rotating}
\usepackage{bm}

\newcommand{\shortname}{ORCA}
\newcommand{\longname}{Over-Reliance-decoupled CAusal multi-task}
\newcommand{\shortNIM}{NDE}

\AtBeginDocument{%
  }

\copyrightyear{2025} 
\acmYear{2025} 
\setcopyright{none}
\acmDOI{}
\acmISBN{}
\renewcommand\footnotetextcopyrightpermission[1]{}

\begin{document}

\title{ORCA: Mitigating Over-Reliance for Multi-Task Dwell Time Prediction with Causal Decoupling}


\author{Huishi Luo}
\email{hsluo2000@buaa.edu.cn}
\orcid{0000-0002-3553-2280}
\author{Fuzhen Zhuang}
\authornote{Corresponding author. Fuzhen Zhuang is also with Zhongguancun Laboratory, Beijing, China.}
\email{zhuangfuzhen@buaa.edu.cn}
\orcid{0000-0001-9170-7009}
\affiliation{%
  \department{Institute of Artificial Intelligence}
  \institution{Beihang University}
  \city{Beijing}
  \country{China}
}

\author{Yongchun Zhu}
\email{zhuyc0204@gmail.com}
\orcid{0000-0002-7096-3190}
\author{Yiqing Wu}
\email{wuyiqing20s@ict.ac.cn}
\orcid{0000-0002-8068-9420}
\affiliation{%
  \institution{Institute of Computing Technology, Chinese Academy of Sciences}
  \city{Beijing}
  \country{China}
}

\author{Bo Kang}
\email{22371112_kblllll@buaa.edu.cn}
\orcid{0009-0000-9757-8113}
\affiliation{%
  \department{Institute of Artificial Intelligence}
  \institution{Beihang University}
  \city{Beijing}
  \country{China}
}

\author{Ruobing Xie}
\orcid{0009-0005-4455-0471}
\email{orcausal@163.com}
\author{Feng Xia}
\email{fengrec@163.com}
\orcid{0009-0002-1408-4535}
\affiliation{%
  \institution{Independent Researcher}
  \city{Beijing}
  \country{China}
}

\author{Deqing Wang}
\email{dqwang@buaa.edu.cn}
\orcid{0000-0001-6441-4390}
\affiliation{%
  \department{SKLSDE, School of Computer Science}
  \institution{Beihang University}
  \city{Beijing}
  \country{China}
}

\author{Jin Dong}
\email{dongjin@baec.org.cn}
\orcid{0009-0006-9708-0220}
\affiliation{%
    \institution{Beijing Advanced Innovation Center for Future Blockchain and Privacy Computing}
  \city{Beijing}
  \country{China}
}
\affiliation{%
  \institution{Beihang University}
  \city{Beijing}
  \country{China}
}

\thanks{This is the authors' preprint version of a paper accepted at CIKM 2025 (short paper). 
The final authenticated version is available at ACM Digital Library: 
\url{https://doi.org/10.1145/3746252.3760898}}

\renewcommand{\shortauthors}{Huishi Luo, Fuzhen Zhuang, Yongchun Zhu, Yiqing Wu, Bo Kang, Ruobing Xie, Feng Xia, Deqing Wang, and Jin Dong}

\begin{abstract}
Dwell time (DT) is a critical post-click metric for evaluating user preference in recommender systems, complementing the traditional click-through rate (CTR). Although multi-task learning is widely adopted to jointly optimize DT and CTR, we observe that multi-task models systematically collapse their DT predictions to the shortest and longest bins, under-predicting the moderate durations. We attribute this \emph{moderate-duration bin under-representation} to over-reliance on the CTR-DT spurious correlation, and propose ORCA to address it with causal-decoupling. Specifically, ORCA explicitly models and subtracts CTR’s negative transfer while preserving its positive transfer. We further introduce (i) feature-level counterfactual intervention, and (ii) a task-interaction module with instance inverse-weighting, weakening CTR-mediated effect and restoring direct DT semantics. ORCA is model-agnostic and easy to deploy. Experiments show an average 10.6\% lift in DT metrics without harming CTR. Code is available \href{https://github.com/Chrissie-Law/ORCA-Mitigating-Over-Reliance-for-Multi-Task-Dwell-Time-Prediction-with-Causal-Decoupling}{here}.
\end{abstract}




\maketitle

\section{Introduction}
\label{sec:intro}

Article recommender systems aim to present users with personalized content that genuinely interests them on online platforms such as Twitter and WeChat. Traditional recommendation methods primarily maximize \textbf{Click-Through Rate (CTR)} as a proxy for user preference~\cite{lin2023autodenoise, luo2025one}. However, CTR’s binary nature provides only coarse feedback and fails to capture the true depth of genuine user engagement. Furthermore, it can be misleading by clickbaits or title–content mismatches in practice~\cite{wang2021clicks, xie2023reweighting}.

\begin{figure*}[!t]
    \centering
    \begin{minipage}{0.7\textwidth}
        \centering
        \includegraphics[height=0.18\textheight, trim=0 5 0 0,clip]{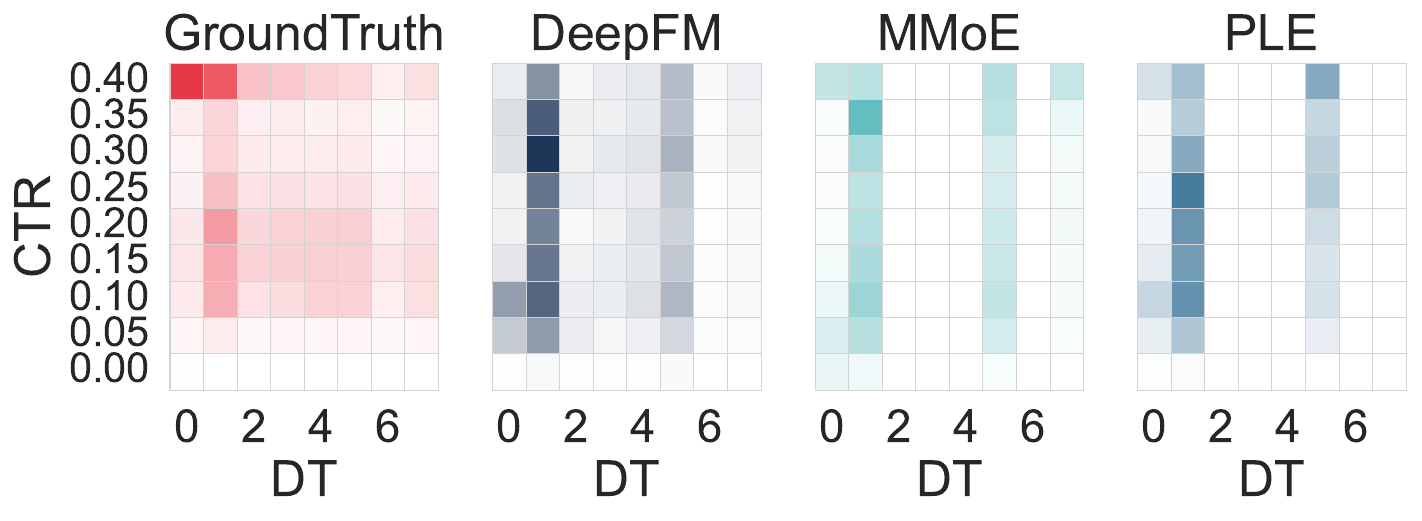}
    \end{minipage}%
    \begin{minipage}{0.2\textwidth}
        \centering
        \includegraphics[height=0.15\textheight, trim=-5 -6 -5 0,clip]{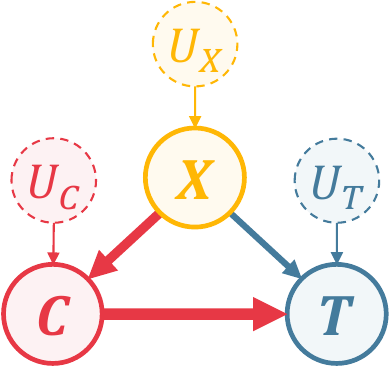}
    \end{minipage}%
    \vspace{-13pt}
    \caption{CTR–DT heatmaps of clicked impressions for the ground truth and classical models, and causal graph of MTL. $X$: input, $C$: clicks, $T$: dwell time, $U$: unobserved factors.}
    \vspace{-5pt}
    \label{fig:intro}
\end{figure*}

Recent studies~\cite{kim2014modeling, zhou2018jump, chen2019follow,xie2023reweighting,zhao2024counteracting, tong2025mindscore} therefore highlight \textbf{Dwell Time (DT)}, the duration a user spends on a clicked item, as a post-click signal that more faithfully reflects depth of user engagement with article content. These works commonly employ multi-task learning (MTL) frameworks~\cite{ma2018mmoe, tang2020ple} that jointly predict CTR and DT. Through knowledge sharing across tasks, MTL enhances both predictions across all tasks, and further mitigating data sparsity of DT labels with abundant CTR supervision. Fig.~\ref{fig:intro} illustrates this process using a causal graph~\cite{pearl2009causality}, where input features $X$ shape both clicks ($C$) and dwell time ($T$), and the sequential path $X \rightarrow C \rightarrow T$ enables knowledge to flow from CTR to DT during optimization.

Despite these advantages, MTL models skew DT predictions toward extremes. Figure~\ref{fig:intro} compares ground-truth CTR–DT distributions over clicked impressions with the predictions of classical models: the single-task DeepFM~\cite{guo2017deepfm}, and the multi-task MMoE~\cite{ma2018mmoe} and PLE~\cite{tang2020ple}. Following common practice~\cite{zhang2023dlm,zhan2022deconfounding,xie2023reweighting}, we discretize continuous DT into eight intervals for classification. We observe that, unlike the ground truth and single-task outputs, MTL predictions collapse onto the extreme bins (0-1, 5, 7), severely under-predicting the moderate bins (2–4, 6) that are appreciable in ground truth. We term this phenomenon \emph{moderate-duration bin under-representation}. It persists across diverse MTL models and degrades the rich continuous DT signal into a coarse, near-binary cue, diminishing its value for assessing nuanced degrees of user interest.

We attribute this issue to the model’s \emph{over-reliance}~\cite{chen2021dependent} on CTR for dwell-time prediction, driven by two main factors. 

\textbf{(1) Task imbalance and optimization bias.} Given the model’s limited representational capacity, the severe label imbalance between CTR and DT supervision ($7.7:1$ in our data) drives the shared layers (over 90 \% of model parameters) to allocate most of their focus on features that benefit CTR prediction at the expense of those critical for DT estimation, strengthening the $X \rightarrow C$ path in Figure~\ref{fig:intro} while weakening $X \rightarrow T$. As a result, the model learns only coarse “short vs. long” distinctions and fails to capture finer patterns that distinguish moderate durations. 

\textbf{(2) Spurious correlations from causal entanglement.} The natural user behavior sequence “impression $\rightarrow$ click $\rightarrow$ read” establishes a strong causal link ($C \rightarrow T$), resulting in high statistical correlation between CTR and DT labels. This high correlation, reinforced by the above-mentioned CTR's dominance during training, encourages the model to adopt CTR features as a predictive \emph{shortcut} for DT. For example, clickbait items with enticing titles often receive high CTR but low DT. The model therefore learns the spurious association “high CTR $\rightarrow$ low DT” and under-predicts dwell time for most high-CTR items, regardless of their actual quality. The reverse shortcut, “low CTR $\rightarrow$ high DT,” likewise emerges, inflating DT estimates for many low-CTR items. 

Together, these factors erode the model’s capacity to learn nuanced dwell-time semantics for moderate-duration.

To address DT’s over-reliance on CTR, we introduce the \textbf{\longname{} (\shortname{})} framework, a causal-decoupling model with the following contributions:

\begin{itemize}
    \item We introduce a negative-dependency extractor that leverages structural causal analysis to explicitly quantify the CTR-mediated effect ($X \rightarrow C \rightarrow T$), subtract its negative component from raw DT predictions, and preserve CTR’s beneficial transfer.

    \item  We propose a feature-level counterfactual masking strategy to weaken the $X \rightarrow C$ path and enhance the direct $X\rightarrow T$ path; and a task-interaction module to expose and mitigate spurious $C \rightarrow T$ shortcuts. By integrating these two mechanisms, \shortname{} mitigate over-reliance on $X \rightarrow C \rightarrow T$ and restore the model’s ability to capture direct $X \rightarrow T$ causal semantics.

    \item \shortname{} is \textbf{easy to deploy} and \textbf{model-agnostic}, requiring only minimal modifications to any MTL backbone. Experiments on two industrial and one public dataset demonstrate an average \textbf{10.6\%} uplift in DT-related metrics without degrading CTR.

\end{itemize}

\section{Related Work}

\subsection{Multi-Task Recommendation}
Multi-task learning (MTL) jointly trains related tasks in recommendation to enable positive transfer~\cite{ song2024mmfi, ma2018esmm,wen2020entire}. 
MMoE~\cite{ma2018mmoe} employs gating mechanisms to control knowledge sharing, PLE~\cite{tang2020ple} separates shared/task-specific parameters, and MetaBalance~\cite{he2022metabalance} adjusts gradients dynamically for task balance. However, these methods mainly promote positive transfer and can even exacerbate CTR-driven over-reliance (Figure~\ref{fig:intro}), whereas \shortname{} explicitly decouples negative transfer. For CTR-DT predictions, Xie et al.~\cite{xie2023reweighting} use DT information to select and reweight click samples. Video watch-time prediction methods~\cite{zhao2024counteracting,zhang2023leveraging,zhan2022deconfounding} tackle a similar problem but assume known total video durations, whereas article DT prediction lacks explicit duration cues.

\subsection{Causal Inference in Recommendation}
Causal inference analyzes cause-effect relations~\cite{pearl2009causality, luo2024survey} and helps reduce biases in recommender systems~\cite{wang2021clicks, ying2023camus, zhang2024adversarial}. Existing works mainly address observed biases, such as popularity~\cite{ zhu2024mitigating} or gender bias~\cite{huang2022achieving}, by treating them as input features. We instead tackle task-level spurious CTR–DT correlations arising from their coupling during joint optimization, rather than from an external attribute. Closest to our work, AECM~\cite{zhang2024adversarial} mitigates CTR–CVR bias in a MTL framework by refining inverse‐propensity weights. However, AECM requires complex adversarial training, while our \shortname{} is a lightweight, plug-and-play causal module compatible with any MTL backbone.

\section{Method}
\label{sec:method}
\subsection{Structural Causal Analysis}
We analyze the causal graph to understand the dwell-time prediction process in an MTL setting. In Figure~\ref{fig:intro}, DT is affected via two pathways:
(i) the direct effect ($X \to T$), which captures genuine user preference; and 
(ii) the CTR-mediated effect ($X \to C\to T$), which represents knowledge transfer but may also introduce negative effect. Due to CTR's stronger supervision, the MTL model focuses on CTR-predictive knowledge, overemphasizing the indirect path and diminishing the features’ direct causal contribution to DT.

To quantify the \emph{total effect} ($TE$) of features on DT estimation, we follow causal effect theory~\cite{pearl2009causality} by comparing an actual state $X = \bm{x}$ and a counterfactual state $X = \bm{x}^*$:

\begin{equation}
\begin{aligned}
    TE
    & = T_{\bm{x},\,C_{\bm{x}}} - T_{\bm{x}^*,\,C_{\bm{x}^*}} \\
    & = f_{T}\bigl(X = \bm{x},\,C = C_{\bm{x}}\bigr) - f_{T}\bigl(X = \bm{x}^*,\,C = C_{\bm{x}^*}\bigr)
\end{aligned}
\end{equation}
The total effect TE can be further decomposed into the \emph{natural indirect effect} ($NIE$), which captures CTR’s mediating influence ($X \to C \to T$), and the \emph{total direct effect} ($TDE$)~\cite{luo2024survey}:

\begin{align}
    NIE
    &= T_{\bm{x}^*,\,C_{\bm{x}}} - T_{\mathbf{x}^*,\,C_{\bm{x}^*}}, 
    \label{eq:nie} \\
    TDE
    &= \mathrm{TE} - \mathrm{NIE} 
    = T_{\bm{x},\,C_{\bm{x}}} - T_{\bm{x}^*,\,C_{\bm{x}}}. 
    \label{eq:tde}
\end{align}
By subtracting NIE from $TE$, $TDE$ effectively removes CTR’s mediating influence on DT estimation and isolates the features' direct causal effect. However, applying $TDE$ directly runs counter to the design intent of MTL, since CTR can also provide positive transfer to DT. This motivates our design of \shortname{}, which selectively removes the negative transfer while preserving positive CTR knowledge.

\subsection{Backbone MTL Model}
We adopt a generic MTL framework, such as PLE~\cite{tang2020ple}, as the base model that predicts the CTR score $\hat{y}^{C}$ and the raw DT logits $\bm{\hat{y}}^{T}$. Specifically, let $\bm{x}$ denote the input feature embeddings. Shared experts and task-specific experts for CTR and DT generate the representations $e^{S}(\bm{x})$, $e^{C}(\bm{x})$, and $e^{T}(\bm{x})$, respectively. These are fed into two independent tower networks, $t^{C}$ and $t^{T}$, producing

\begin{equation}
\label{eq:ycyt}
\hat{y}^{C}(\bm{x}) = t^{C}\left([e^S(\bm{x}), e^C(\bm{x})]\right), \quad
\bm{\hat{y}}^{T}(\bm{x}) = t^{T}\left([e^S(\bm{x}), e^T(\bm{x})]\right).
\end{equation}
In industrial recommender systems, decision logic focuses on the relative level of dwell time rather than its exact value. Therefore, following prior work~\cite{zhang2023dlm, zhan2022deconfounding}, we discretise the continuous DT signal into $M$ non-overlapping intervals for classification.

\begin{figure*}[t!]
\centering
\includegraphics[height=0.35\textheight, trim=0 0 0 0, clip]{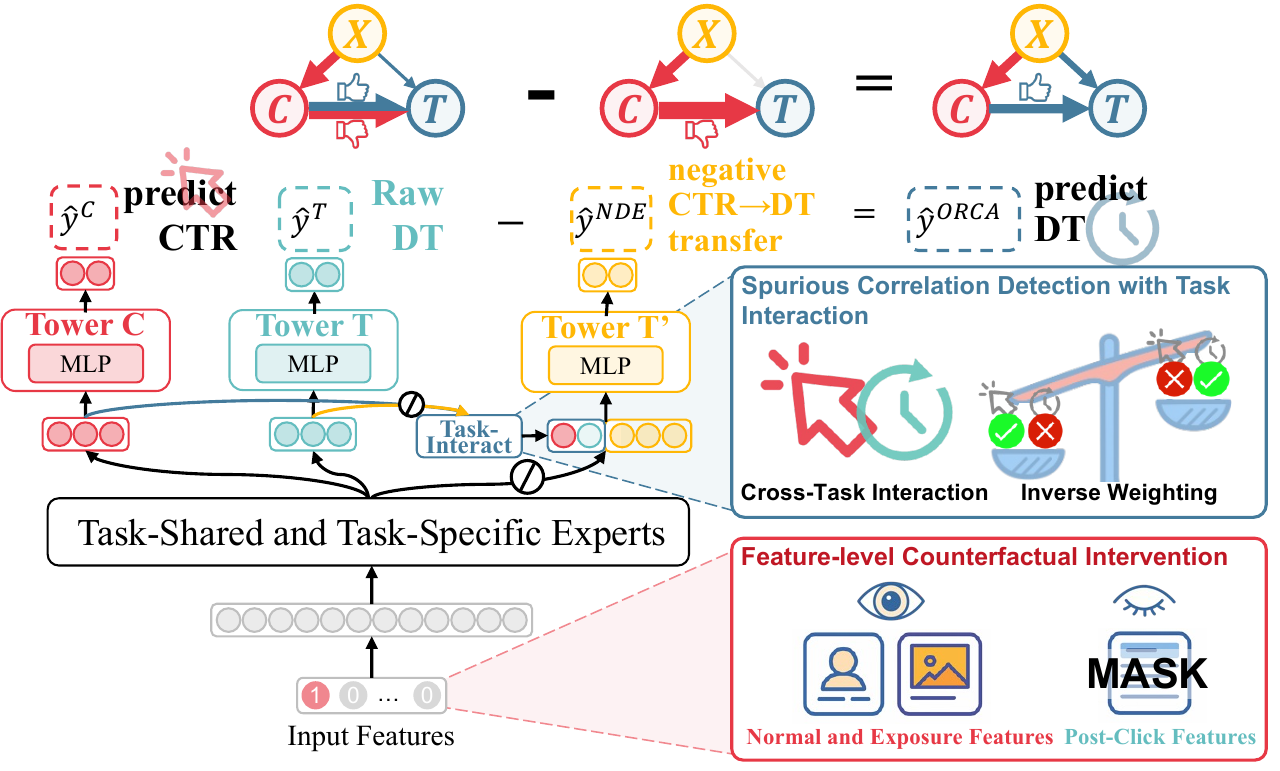}
\vspace{-8pt}
\caption{Overview of ORCA. Negative Dependency Extractor ($t^{T'}$) explicitly models the model’s over-reliance on CTR.}
\vspace{-5pt}
\label{fig:model}
\end{figure*}

\subsection{Negative CTR Dependency Extractor (NDE)}
\label{sec:nde}

Inspired by $TDE$ (Equation~\ref{eq:tde}), we \emph{explicitly} model the negative dependency of DT on CTR and subtract it from the original DT prediction. We augment the MTL backbone with an auxiliary tower $t^{T'}$, which directly estimates the CTR-induced negative effect $\bm{\hat{y}}^{\shortNIM{}}$, as in Figure~\ref{fig:model}. Its input is the shared-expert representation $e^{S}(\bm{x})$, since CTR dominance in $\bm{\hat{y}}^{T}$ mainly arises from shared experts, whose parameters are dominated by CTR supervision during training.

\begin{equation}
\label{eq:yB}
\bm{\hat{y}}^{\shortNIM{}}(\bm{x}) = t^{T'}\left([e^{S}(\bm{x})]\right).
\end{equation}
The refined DT prediction is given by:

\begin{equation}
\label{eq:yF}
\bm{\hat{y}}^{\shortname{}}(\bm{x}) = \bm{\hat{y}}^{T}(\bm{x}) - \bm{\hat{y}}^{\shortNIM{}}(\bm{x}).
\end{equation}
Higher $\bm{\hat{y}}^{\shortNIM{}}(\bm{x})$ values indicate stronger CTR-induced bias. By subtracting them from $\bm{\hat{y}}^{T}(\bm{x})$, we mitigate over‐dependence on CTR and produce debiased logits for the redistributed DT probability.

\subsection{Feature-level Counterfactual Intervention}
\label{sec:fea}

Building on $TDE$ (Equation~\ref{eq:tde}), \shortNIM{} aims to approximate the counterfactual term $T_{\bm{x}^*,\,C_{\bm{x}}}$ by blocking the features’ direct effect via $X=\bm{x}^*$ so that any causal influence on DT passes through the CTR mediator $C$. Unfortunately, in MTL the CTR–mediated effect is implicitly embedded in shared representations rather than exposed as a controllable input factor, making it difficult to intervene as one would for known biases (e.g., gender). In other words, the direct path ($X \to T$) and the indirect path ($X \to C \to T$) are entangled, making rigorous counterfactual estimation challenging.

We address this by intervening on input features to induce $t^{T'}$ to disentangle CTR effects from shared representations. Using “leave-one-feature-out” analysis \cite{liu2013leave,nakanishi2024evolving}, we identify features that almost exclusively affecting DT (e.g., article content, post-click keywords), denoted as \emph{post-click} features $x_{\text{post}}$. Intuitively, $x_{\text{post}}$ conveys the direct causal signal $X \rightarrow T$. Consequently, during \shortNIM{} training, we mask them with probability $p_{\text{fea}}$:

\begin{equation}
\bm{x} = \left(x \setminus x_{\text{post}}, [\mathrm{MASK}]\right).
\label{eq:fea}
\end{equation}
Masking $x_{\text{post}}$ drives \shortNIM{} to capture CTR-mediated cues; subtracting $\hat{\bm y}^{\shortNIM{}}$ then reduces the final predictor’s reliance on these cues and promotes use of features along the direct $X \to T$ path.

\subsection{Spurious Correlation Detection with Task Interaction (SCD)}
\label{sec:inter}

As mentioned in Section \ref{sec:intro}, DT predictions in conventional MTL are heavily biased toward the extreme (very short or very long) bins, while the moderate-duration buckets are severely underrepresented. We attribute this distributional imbalance to \emph{DT’s over-reliance on the direct CTR signal as a shortcut} to predict engagement. This shortcut captures superficial CTR–DT correlations, such as “high click $\rightarrow$ short dwell” or “no / low click $\rightarrow$ long dwell”, and inevitably leads to a loss of semantic capacity for discriminating moderate dwell times. To correct such spurious correlations, we explicitly model cross-task CTR–DT relationships by introducing a cross-interaction module that fuses the task-specific representations of CTR and DT before feeding them into $t^{T'}$:

\begin{equation}
\label{eq:inter}
\bm{\hat{y}}^{\shortNIM{}}(\bm{x}) = t^{T'}\left([e^{S}(\bm{x}), \mathrm{Inter}(e^{C}, e^{T})]\right).
\end{equation}
where $\mathrm{Inter}(\cdot,\cdot)$ denotes a feature interaction function. In this work, we adopt multi-head self-attention~\cite{song2019autoint} due to its effective performance.

Furthermore, to preserve beneficial knowledge transfer from CTR while selectively mitigating negative transfer, \shortNIM{} must distinguish the negative impact within the cross-task interaction. An intuitive hypothesis is that instances with accurate CTR but inaccurate DT predictions are disproportionately influenced by spurious correlations. Thus, we assign higher weights to their \shortNIM{} loss. Specifically, the instance weight is computed as the product of the raw DT loss and the inverse of the CTR loss, i.e.,

\begin{equation}
\label{eq:w_loss}
w_{i}=\left(\mathcal{L}_{C, i}\right)^{-\alpha} \cdot \mathcal{L}_{T, i}+ \gamma, \alpha \geq 0,
\end{equation}
where $\alpha$ is a coefficient for the CTR loss and $\gamma$ is an offset. Samples with low CTR loss (i.e., accurate CTR predictions) and high DT loss receive higher weights, thus recalibrating \shortNIM{}'s focus on correcting spurious negative correlations from CTR to DT.

From a causal perspective, the cross-task interaction module explicitly instantiates the $C \to T$ pathway in the causal graph. From the perspective of MTL optimization, instances exhibiting a significant CTR–DT relationship are amplified in the interaction representation. Furthermore, among these, instances reflecting detrimental correlations are prioritized via loss weighting. This enables the model to calibrate the final DT predictions and recover the semantic capacity for moderate dwell times, which are otherwise underrepresented due to spurious correlations.

\subsection{Optimization Objectives}
\label{subsec:optim}

\shortname{} uses a multi-task learning approach by minimizing the cross-entropy losses for predictions $\hat{y}^{C}(\bm{x})$, $\hat{y}^{T}(\bm{x})$, and $\hat{y}^{\shortname{}}(\bm{x})$:

\begin{equation}
\mathcal{L}=\mathcal{L}_{C}+\mathcal{L}_{T}+\mathcal{L}_{\shortname{}},
\label{eq:loss}
\end{equation}
where the DT-related losses $\mathcal{L}_{T}$ and $\mathcal{L}_{\shortname{}}$ are computed for clicked instances only. According to Section~\ref{sec:inter}, 


\begin{equation}
\mathcal{L}_{\shortname{}}
=\frac{1}{\sum_{i \in {\mathcal{S}}^{+}} w_{i}} \sum_{i \in {\mathcal{S}}^{+}}\left(-w_{i} \sum_{j=1}^{M} y_{i, j}^{T} \log \left(\widehat{y}_{i, j}^{\shortname{}}(x)\right)\right),
\end{equation}
where $w_{i}$ are sample weights from Equation~(\ref{eq:w_loss}) for clicked samples ${\mathcal{S}}^{+}$, $y_i^{T}\in \mathbb{R}^M$ is the ground-truth DT vector, and $M$ is the number of discretized DT intervals. To prevent interference between the backbone and the bias-capturing \shortNIM{} module, we stop gradients flowing from \shortNIM{} into the backbone input representations.

\section{Experiments}
\label{sec:experiment}

\subsection{Experimental Settings}
\label{sec:setting}
\subsubsection{Datasets}
We collect impression and reading logs from a real-world article recommendation system, forming DS183K and DS651K. After performing a preliminary “leave-one-feature-out” analysis~\cite{liu2013leave,nakanishi2024evolving}, we identify \texttt{publication timestamp} and \texttt{category} as the post-click features for counterfactual intervention (Section~\ref{sec:fea}). Dwell time is discretized into bins to align with the system's online setting~\cite{zhang2023dlm, zhan2022deconfounding}. System details will be disclosed later due to anonymity. For reproducibility, we also use the public Tenrec dataset~\cite{yuan2022tenrec}, which records various user interactions but does not include the item features needed for intervention. The dataset statistics are as follows: DS183K (183K, 13K users, 77K articles, 25K clicks), DS651K (651K, 24K users, 166K articles, 84K clicks), and Tenrec (9.8M, 66K users, 71K articles, 1.57M clicks).

\begin{table*}[!t]
  \renewcommand\arraystretch{0.8}
  \centering

  \caption{Results on three datasets. * indicates $p<0.05$ based on a paired t-test between our method and its backbone. Note that $\downarrow$MAE\textsubscript{class} and $\downarrow$RMSE\textsubscript{class} are the primary DT metrics in our task setting.}
  \vspace{-5pt}
  \setlength{\tabcolsep}{6pt}{
    \begin{tabular}{c|l|ccccccc|c}
    \toprule    
    \multirow{16}[10]{*}{\begin{sideways}\textbf{DS183K}\end{sideways}} & \multicolumn{1}{c|}{
    \multirow{2}[4]{*}{\textbf{Methods}}} & \multicolumn{7}{c|}{\textbf{DT}} & \textbf{CTR}  \\ 
\cmidrule(r){3-9} \cmidrule{10-10}          &       & \textbf{MAE\textsubscript{class}} & \textbf{RMSE\textsubscript{class}} & \textbf{F1} & \textbf{Precision} & \textbf{Recall} & \textbf{MAE} & \textbf{RMSE} & \textbf{AUC} \\
\cmidrule{2-10}          & NFM   & 1.427  & 2.286  & 0.407  & 0.201  & 0.253  & 57.56  & 108.19  & 0.699  \\
          & DeepFM & 1.308  & 2.123  & 0.431  & 0.244  & 0.274  & 56.41  & 103.89  & 0.725  \\
          & AutoInt & 1.328  & 2.149  & 0.425  & 0.231  & 0.270  & 56.89  & 104.48  & 0.734  \\
          & AFN   & 1.445  & 2.310  & 0.407  & 0.210  & 0.254  & 58.64  & 110.10  & 0.704  \\
          & DCN-V2 & 1.415  & 2.254  & 0.417  & 0.243  & 0.254  & 58.15  & 107.63  & 0.730  \\
\cmidrule{2-10}          & ESMM  & 1.527  & 2.439  & 0.387  & 0.158  & 0.237  & 59.68  & 114.98  & 0.642  \\
          & AITM  & 1.441  & 2.326  & 0.407  & 0.234  & 0.255  & 58.56  & 110.72  & 0.712  \\
          & EXUM  & 1.431  & 2.237  & 0.411  & 0.225  & 0.265  & 64.03  & 111.61  & 0.697  \\
\cmidrule{2-10}
          & MMoE  & 1.452  & 2.343  & 0.403  & 0.195  & 0.250  & 58.29  & 110.65  & 0.737  \\
          & \ +\shortname{}  & \textbf{1.179*}  & \textbf{1.989*}  & 0.448*  & \textbf{0.300*}  & 0.295*  & \textbf{54.91*}  & \textbf{102.63*}  & 0.739  \\
\cmidrule{2-10}
          & PLE   & 1.384  & 2.248  & 0.416  & 0.218  & 0.266  & 57.71  & 108.09  & \textbf{0.739}  \\
          & \ +\shortname{}  & 1.201*  & 2.017*  & 0.444*  & 0.278*  & 0.287*  & 55.42*  & 105.94*  & 0.739  \\
\cmidrule{2-10}
          & MetaBalance & 1.334  & 2.170  & 0.422  & 0.214  & 0.274  & 58.19  & 106.05  & 0.738  \\
          & \ +\shortname{}  & 1.249*  & 2.113*  & \textbf{0.471*}  & 0.224*  & \textbf{0.302*}  & 56.62  & 103.73*  & 0.737  \\
    \midrule
    \multirow{16}[10]{*}{\begin{sideways}\textbf{DS651K}\end{sideways}} & \multicolumn{1}{c|}{
    \multirow{2}[4]{*}{\textbf{Methods}}} & \multicolumn{7}{c|}{\textbf{DT}} & \textbf{CTR}  \\ 
\cmidrule(r){3-9} \cmidrule{10-10}        &       & \textbf{MAE\textsubscript{class}} & \textbf{RMSE\textsubscript{class}} & \textbf{F1} & \textbf{Precision} & \textbf{Recall} & \textbf{MAE} & \textbf{RMSE} & \textbf{AUC} \\
\cmidrule{2-10}          & NFM   & 1.551  & 2.287  & 0.315  & 0.268  & 0.278  & 74.50  & 119.89  & 0.746  \\
          & DeepFM & 1.492  & 2.237  & 0.337  & 0.293  & 0.295  & 72.79  & 119.14  & 0.759  \\
          & AutoInt & 1.527  & 2.266  & 0.324  & 0.274  & 0.285  & 73.51  & 118.59  & 0.763  \\
          & AFN   & 1.577  & 2.329  & 0.308  & 0.256  & 0.268  & 74.12  & 119.95  & 0.749  \\
          & DCN-V2 & 1.524  & 2.252  & 0.331  & 0.278  & 0.281  & 73.51  & 118.90  & 0.760  \\
\cmidrule{2-10}          & ESMM  & 1.503  & 2.253  & 0.322  & 0.257  & 0.287  & 71.74  & 114.83  & 0.727  \\
          & AITM  & 1.520  & 2.284  & 0.316  & 0.266  & 0.282  & 71.72  & 116.64  & 0.739  \\
          & EXUM  & 1.528  & 2.269  & 0.313  & 0.260  & 0.277  & 71.26  & 113.64  & 0.744  \\
\cmidrule{2-10}
          & MMoE  & 1.497  & 2.261  & 0.327  & 0.279  & 0.286  & 70.97  & 116.08  & 0.755  \\
          & \ +\shortname{}  & 1.398*  & \textbf{1.999*}  & \textbf{0.361*}  & 0.323*  & 0.310*  & \textbf{68.29*}  & 110.80*  & 0.759  \\
\cmidrule{2-10}
          & PLE   & 1.501  & 2.269  & 0.324  & 0.270  & 0.288  & 69.98  & 115.44  & 0.763  \\
          & \ +\shortname{}  & 1.414*  & 2.128*  & 0.356*  & \textbf{0.325*}  & 0.304*  & 69.43  & 112.37*  & 0.764  \\
\cmidrule{2-10}
          & MetaBalance & 1.491  & 2.254  & 0.324  & 0.264  & 0.286  & 69.10  & 113.75  & \textbf{0.765}  \\
          & \ +\shortname{}  & \textbf{1.366*}  & 2.061*  & 0.361*  & 0.317*  & \textbf{0.327*}  & 68.60  & \textbf{108.59*}  & 0.763  \\
    \midrule

    \multirow{8}[10]{*}{\begin{sideways}\textbf{Tenrec}\end{sideways}} & \multicolumn{1}{c|}{
    \multirow{2}[4]{*}{\textbf{Methods}}} & \multicolumn{7}{c|}{\textbf{DT}} & \textbf{CTR}  \\ 
\cmidrule(r){3-9} \cmidrule{10-10}         &       & \textbf{MAE\textsubscript{class}} & \textbf{RMSE\textsubscript{class}} & \textbf{F1} & \textbf{Precision} & \textbf{Recall} & \textbf{MAE} & \textbf{RMSE} & \textbf{AUC} \\
\cmidrule{2-10}          & DeepFM & 1.045  & 1.645  & 0.375  & 0.220  & 0.220  & 39.00  & 63.80  & 0.958  \\
          & AutoInt & 1.045  & 1.641  & 0.371  & 0.215  & 0.215  & 38.86  & 63.31  & 0.958  \\
\cmidrule{2-10}          & MMoE  & 1.044  & 1.650  & 0.376  & 0.226  & 0.226  & 38.80  & 64.32  & 0.958  \\
          & \ +\shortname{}  & 1.018*  & 1.505*  & 0.392*  & \textbf{0.343*}  & 0.257*  & 38.11  & 63.77  & 0.957  \\
\cmidrule{2-10}
          & PLE   & 1.041  & 1.646  & 0.379  & 0.230  & 0.230  & 38.79  & 64.41  & 0.956  \\
          & \ +\shortname{}  & \textbf{0.964*}  & \textbf{1.473*}  & \textbf{0.403*}  & 0.338*  & \textbf{0.259*}  & \textbf{37.36*}  & \textbf{62.28*}  & \textbf{0.958}  \\
          
    \bottomrule
    \end{tabular}}%
  \label{tab:result}%
  \vspace{-5pt}
\end{table*}%

\subsubsection{Baselines}
We compare \shortname{} against a range of representative single-task and multi-task models. For single-task objectives focusing only on CTR or DT, we implement classical models such as \textbf{NFM}~\cite{he2017neural}, \textbf{DeepFM}~\cite{guo2017deepfm}, \textbf{AutoInt}~\cite{song2019autoint}, \textbf{AFN}~\cite{cheng2020afn}, and \textbf{DCN-V2}~\cite{wang2021dcn}. For multi-task models, we benchmark against \textbf{MMoE}~\cite{ma2018mmoe}, \textbf{ESMM} \cite{ma2018esmm}, \textbf{PLE}~\cite{tang2020ple}, \textbf{AITM}~\cite{xi2021modeling}, \textbf{MetaBalance}~\cite{he2022metabalance}, and \textbf{EXUM}~\cite{wu2025exum} serving both as baselines and potential base models for \shortname{}.

\subsubsection{Implementation}
To test \shortname{}’s generality, we apply it to MMoE, PLE, and MetaBalance (MB) backbones. Hyper-parameters are tuned via grid search. Continuous dwell time is discretized via equidistant binning on $\log t$ distributions and modeled as a multi-class task, as the log transformation approximates a Gaussian distribution in practice~\cite{zhou2018jump, xie2023reweighting}. We split $\log t$ into eight intervals, aligning with prior work \cite{wu2022quality, xie2023reweighting, zhang2023dlm, zhan2022deconfounding} and our online system.

\subsubsection{Metrics}
(1) For DT, we report both regression and classification metrics. For regression, we report MAE and RMSE~\cite{chen2019follow, ma2019knowing}, assigning each predicted bin its median value for comparison with the continuous ground truth. Since DT is modeled as a multi-class task, we also report Weighted F1-score, Macro Precision, Macro Recall, and class-based MAE\textsubscript{class} and RMSE\textsubscript{class}. We prioritize classification over regression metrics, since class-based evaluation better aligns with industrial system requirements. Regression metrics are reported for reference but can be biased due to non-uniform binning. (2) For CTR prediction, we use AUC~\cite{guo2017deepfm, ma2018mmoe}, but our primary focus is on improving DT performance.

\subsection{Model Analyses}


\subsubsection{Effectiveness and Universality of \shortname{}} 
We implement \shortname{} on various MTL backbones and compare it against both single- and multi-task baselines to validate its effectiveness and universality. Table~\ref{tab:result} shows: (1) \shortname{} outperforms all baselines on DT metrics with an average 10.6\% uplift. (2) \shortname{}'s CTR performance matches the best baseline, proving negative-transfer modeling does not hurt CTR. (3) \shortname{} delivers consistent gains on MMoE, PLE, and MetaBalance backbones, demonstrating robustness and universality.

\subsubsection{Ablation Study}
\begin{figure*}[!t]
    \centering
    \begin{minipage}{0.25\textwidth}
        \centering
        \includegraphics[height=0.120\textheight, trim=6 5 6 0,clip]{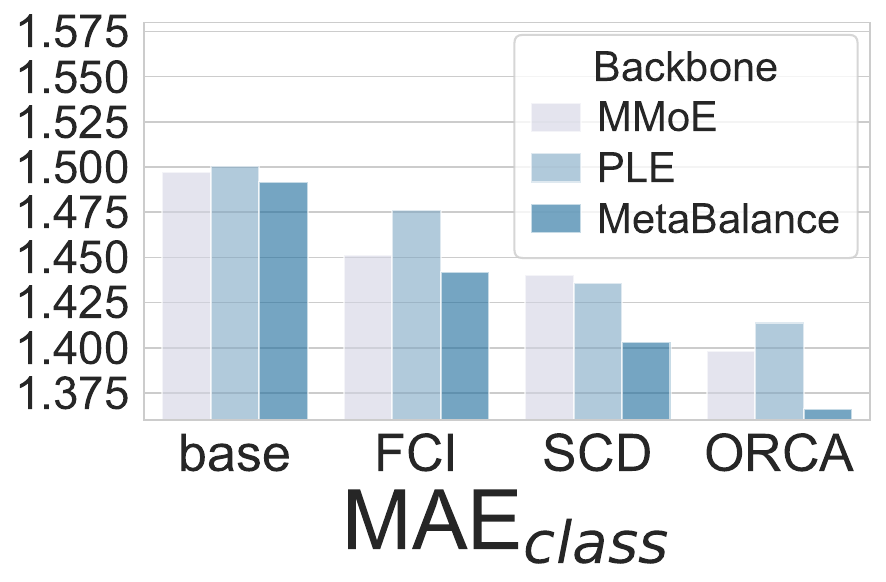}
    \end{minipage}%
    \begin{minipage}{0.25\textwidth}
        \centering
        \includegraphics[height=0.120\textheight, trim=6 5 6 0,clip]{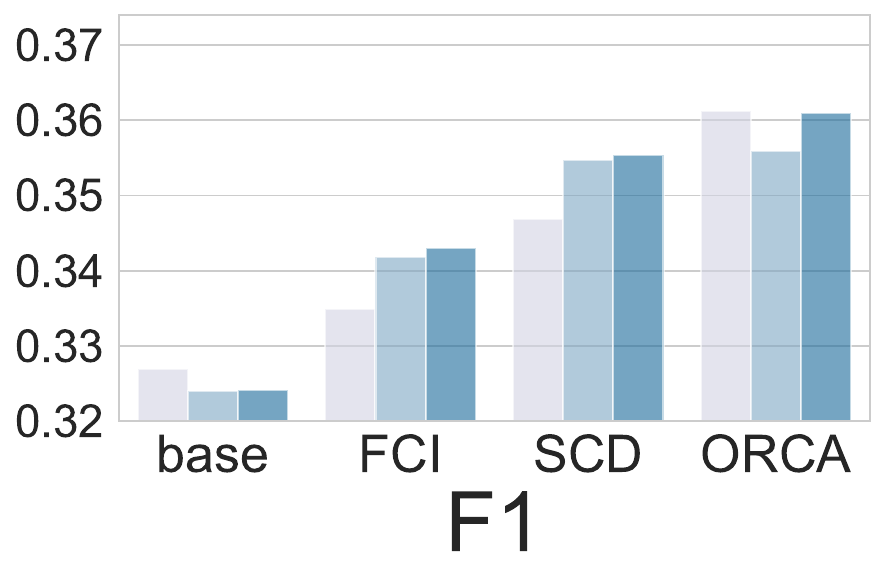}
    \end{minipage}%
    \begin{minipage}{0.25\textwidth}
        \centering
        \includegraphics[height=0.120\textheight, trim=6 5 6 0, clip]{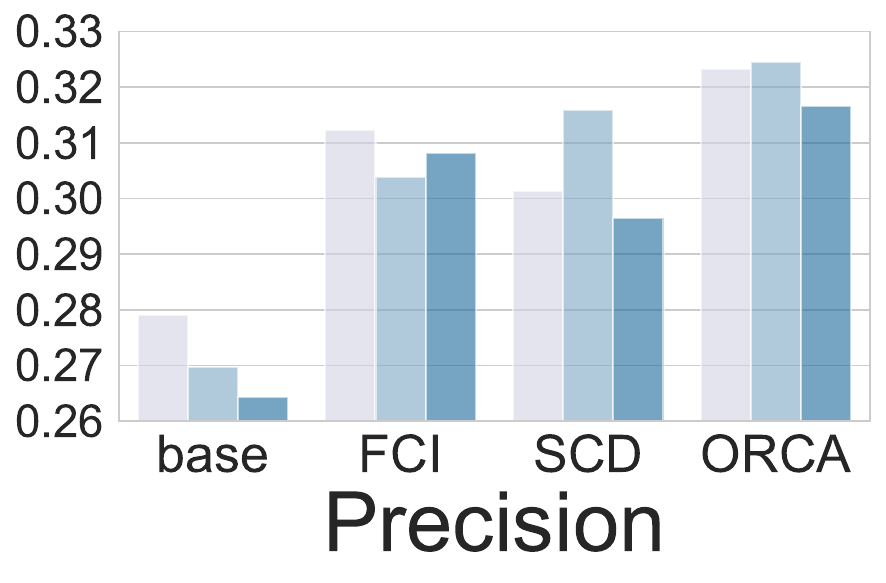}
    \end{minipage}%
    \vspace{-5pt}
    \caption{Ablation study on DS651K.}
    \label{fig:ablation_3day}
\end{figure*}

We assess each \shortname{} component on DS651K via four variants: (1) \textbf{base}: backbone only; (2) \textbf{FCI}: add the auxiliary tower $t^{T'}$ (Section~\ref{sec:nde}) and apply Feature-level Counterfactual Intervention (Section~\ref{sec:fea}); (3) \textbf{SCD}: use task interaction representation and inverse sample weighting (Section~\ref{sec:inter}); (4) \shortname{}: all components combined. Figure~\ref{fig:ablation_3day} shows that each variant significantly outperforms its backbone on DT metrics ($p<0.05$).

\subsubsection{Case Study: Restoring Direct Predictive Ability for DT}
\begin{figure*}[!t]
    \centering
    \begin{minipage}{0.25\textwidth}
        \centering
        \includegraphics[height=0.15\textheight, trim=7 9 6 6.2,clip]{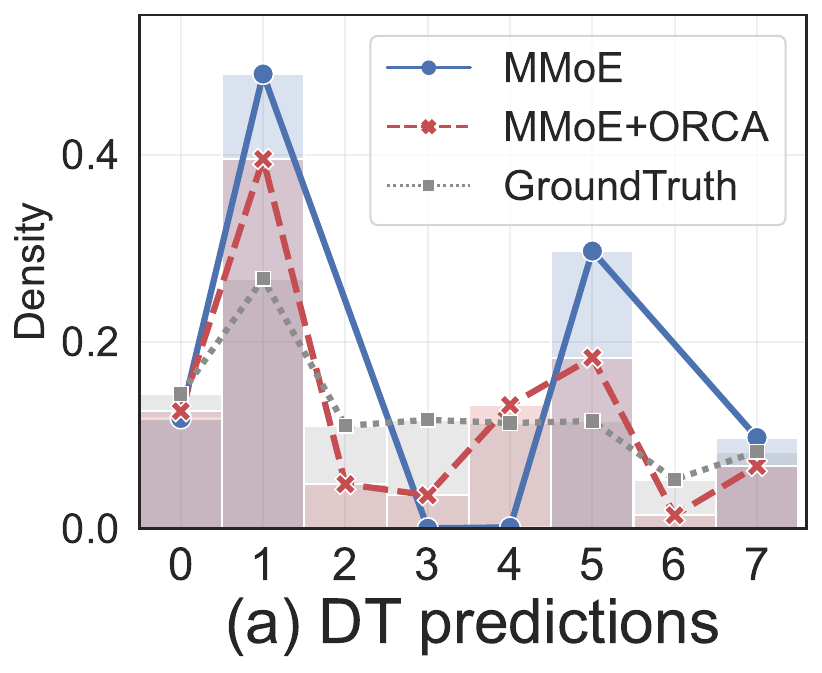}
    \end{minipage}%
    \begin{minipage}{0.28\textwidth}
        \centering
        \includegraphics[height=0.15\textheight, trim=0 9 6 2, clip]{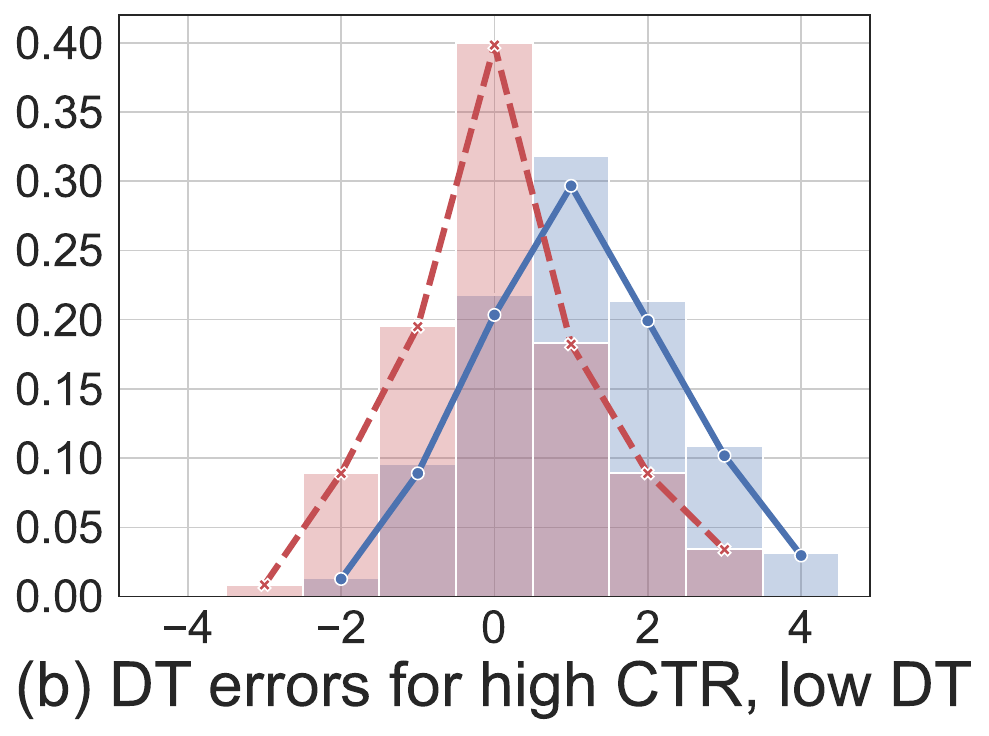}
    \end{minipage}%
    \begin{minipage}{0.25\textwidth}
        \centering
        \includegraphics[height=0.15\textheight, trim=8 9 6 8,clip]{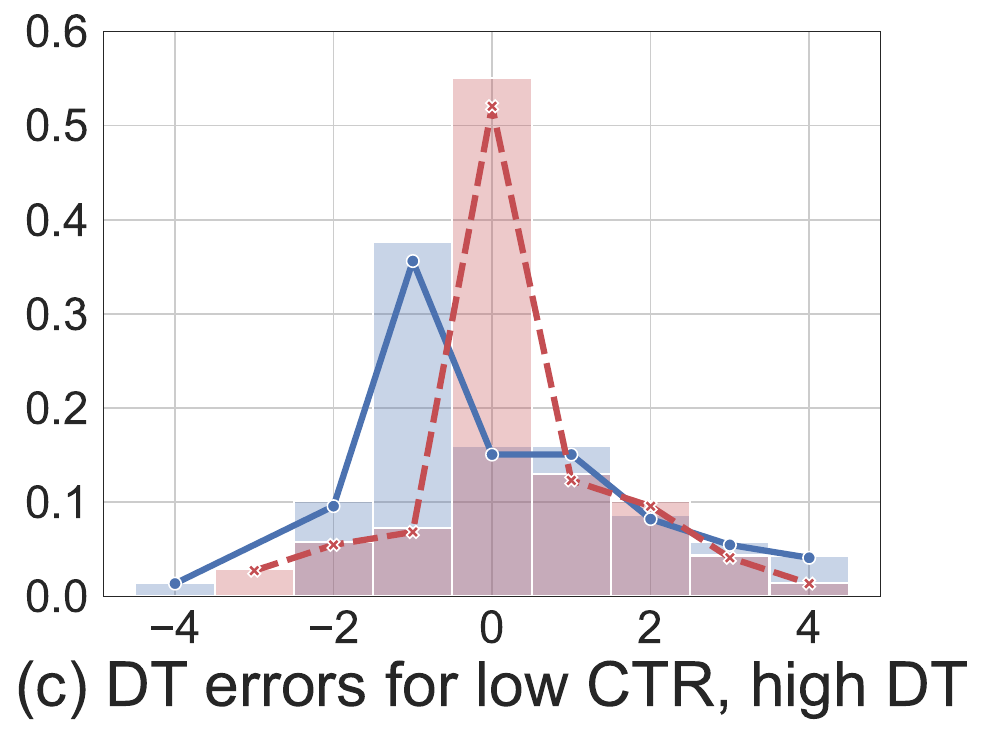}
    \end{minipage}
    
    \vspace{-5pt}
    \caption{Moderate DT recovery and over-reliance mitigation.}
    \vspace{-5pt}
    \label{fig:analysis}
\end{figure*}

We illustrate \shortname{}’s impact on DT prediction by visualizing the distributional differences between the baseline MMoE and \shortname{} on the DS651K dataset. As shown in Figure~\ref{fig:analysis}(a), MMoE predicts almost no samples in bins 2, 3, 4, and 6, despite each of these bins having a modest presence in the ground truth. In contrast, \shortname{} allocates a noticeable share of its predictions to these duration bins, indicating that \shortname{} has restored the model’s ability to represent moderate DT semantics. To further quantify how \shortname{} mitigates DT’s over-reliance on CTR, we analyze prediction errors for articles with high CTR but low DT, and vice versa. Figures~\ref{fig:analysis}(b) and (c) show that MMoE exhibits a bias positively correlated with CTR, overestimating DT for high-CTR articles and underestimating DT for low-CTR articles. In contrast, \shortname{} yields nearly unbiased errors, demonstrating its capability to correct CTR-induced biases.

\section{Conclusion and Future Work}
In this work, we propose the \shortname{} framework to alleviate dwell time prediction’s over-reliance on CTR in multi-task learning. \shortname{} calibrates the DT distribution by explicitly modeling the negative dependency bias. It is a simple yet effective model-agnostic framework. Experiments show that \shortname{} successfully restores beneficial DT semantic capacity without compromising CTR performance. In future work, we will investigate whether a similar over-reliance issue arises in the CTR–CVR problem and explore the applicability of \shortname{} for that task.

\section*{GenAI Usage Disclosure}

We used OpenAI’s ChatGPT to review and refine English grammar and style, and GitHub Copilot to complete repetitive code. Any text or code generated or suggested by these tools was reviewed and finalized by the authors. The authors take full responsibility for the accuracy and originality of all material in this work.

\begin{acks}
Supported by the National Key Research and Development Program of China under Grant No. 2024YFF0729003, the National Natural Science Foundation of China under Grant Nos. 62176014 and 62276015, and the Fundamental Research Funds for the Central Universities.
\end{acks}

\bibliographystyle{ACM-Reference-Format}
\bibliography{ref}


\end{document}